\documentstyle[preprint,pre,aps]{revtex}

\begin{document}

\title{\bf The Reaction-Diffusion Front
for $A+B \rightarrow\emptyset$ in One Dimension.}
\author{G.T. Barkema}
\address{Institute for Advanced Study, Olden Lane, Princeton, NJ
08540, USA.} 
\author{M.J. Howard and J.L. Cardy}
\address{Department of Physics, Theoretical Physics, University of
Oxford, \\ 1 Keble Road, Oxford, OX1 3NP, United Kingdom.}

\maketitle

\begin{abstract}
We study theoretically and numerically the steady state
diffusion controlled reaction $A+B\rightarrow\emptyset$,
where currents $J$ of $A$ and $B$ particles are
applied at opposite boundaries. For a reaction rate
$\lambda$, and equal diffusion constants $D$, we find that when
$\lambda J^{-1/2} D^{-1/2}\ll 1$ the reaction front is well
described by mean field theory. However, for $\lambda
J^{-1/2} D^{-1/2}\gg 1$, the front acquires a Gaussian profile -
a result of noise induced wandering of the reaction front
center. We make a theoretical prediction for this profile which is in
good agreement with simulation. Finally, we investigate the intrinsic
(non-wandering) front width and find results consistent with scaling
and field theoretic predictions.
\end{abstract}



Recently there has been considerable interest in the properties of
diffusion limited chemical reactions \cite{KK,OTB}. Processes such as
$A+B\rightarrow\emptyset$, where diffusing chemicals react
irreversibly are believed to have many applications in physical,
chemical, and biological systems. Particular attention has been paid
to cases where a reaction front is formed between regions dominated by
$A$ or $B$ particles. Such a situation can arise in the case where the
two species are initially entirely segregated
\cite{GR,THKTW,CDC,AHLS1,AHLS3,AHLS2,LC,C}, or
alternatively, and more simply, in
a steady state situation, where $A$ and $B$ particles are
injected at equal rates at opposite boundaries
\cite{LC,BR,CD,HC}. In 
this letter we study the latter model, in the case of equal diffusion
constants $D$ for the two species.
The simplest description of these systems is provided by the
inhomogeneous mean-field rate equations for the particle densities $a$
and $b$, where it is assumed that the reaction rate
$R=\lambda ab$:
\begin{eqnarray}
& {\partial a \over\partial t}=D\nabla^2 a-\lambda ab \label{mf1} \\
& {\partial b \over\partial t}=D\nabla^2 b-\lambda ab. \label{mf2}
\end{eqnarray}
These equations predict a reaction front width which scales as
$w_{mf}\sim(\lambda J/D^2)^{-1/3}$, where $J$ are the (equal) imposed
currents of $A$ and $B$ particles at the boundaries. However, it is
well known that below a critical spatial dimension $d_c=2$
\cite{LC,CD,HC} 
microscopic density fluctuations become relevant, and as a result the
mean field approach breaks down. For $d=1$ Cornell and Droz
\cite{CD} have suggested that the fluctuations modify the scaling of
the width to $w\sim (J/D)^{-1/2}$. Numerical simulations \cite{C,CD}
have 
broadly confirmed these conclusions. However, there has been recent
controversy in the time dependent version of the model, with initially
separated reactants, over the existence of multiscaling in the spatial
moments of the one dimensional reaction front \cite{AHLS2,C}. More
recent simulations by Cornell \cite{C} have also indicated that
the one dimensional profile is accurately described by a
Gaussian. Hitherto this result has not been understood.

Theoretical approaches to understanding the crucial role played by
fluctuations have centered on mappings of the microscopic
dynamics of the reaction-diffusion system onto a quantum field
theory \cite{LC,HC,L}. This has allowed the effects of fluctuations to
be systematically included by summing sets of Feynman diagrams.
Renormalization group (RG) techniques have then been employed to form
a perturbation expansion in $\epsilon=2-d$. These calculations have
confirmed the modified scaling in $d=1$, as well as pointing to the
existence of power law tails, both in the densities and in the
reaction front:
\begin{eqnarray}
& & R=AD(J/D)^{19\over
12}|x|^{(9-7\epsilon)/12}\exp{\left({-B(J/D)^{1\over 2} 
|x|^{(3-\epsilon)/2}}\right)} \nonumber
\\ & & \qquad\quad\qquad +CD^2J^{-1}|x|^{-7+2\epsilon} +\ldots
\end{eqnarray}
Here $A,B,C$ are universal dimension dependent constants. One
consequence of the power law tails is that sufficiently 
high order spatial moments of an evolving time dependent reaction
front (with initially segregated reactants) should exhibit
multiscaling. These results were derived on the basis of a
finite reaction rate, which under the RG was found to flow to a
universal $O(\epsilon)$ fixed point. However, previous simulations of
one dimensional reaction-diffusion systems have employed an infinite
reaction rate, which enforces complete segregation between the two
species. Furthermore only single occupancy of a given lattice
site has previously been permitted. In this letter, we relax these
restrictions by simulating a system with both multiple site occupancy
and an adjustable, finite, reaction rate $\lambda$. This
model is closer to that used in the analytic RG calculations.


Our model consists of a one dimensional lattice with $L$ sites, on
which particles of types $A$ and $B$ are located. In addition to this
the model features reservoirs containing either $A$ or $B$
particles. The total number of particles of each species
was set equal: $N_A=N_B=N/2$. Three distinct processes
take place:

(1) $A$ and $B$ particles located on the lattice hop to neighboring
sites in each direction with a hopping rate $h$, which we set equal to
1 (corresponding to $D=1$ in the continuum theory).

(2) Each $A$ particle can react with each $B$ particle on the same
lattice site, with a reaction rate $\lambda$.
After each reaction both particles are removed from the
lattice and placed in their respective reservoirs.

(3) Each $A$ ($B$) particle in the reservoir is inserted onto the
leftmost (rightmost) site in the lattice with an insertion rate
$i$. Clearly 
$J=N_{res}\cdot i$, where $N_{res}$ is the number of particles in the
reservoir. The purpose of these reservoirs is to break up correlations
between particle annihilation inside the reaction front and particle
reinsertion at the boundaries - the larger the reservoirs the smaller
the correlations. The same effect can be achieved by
increasing the system size, but this is computationally far less
efficient, as in that case particles have to hop large distances
before they can reach the reaction zone.


We carry out the simulations with rare-event dynamics (RED).
In this approach no fixed time increment is present. First, in a
specific configuration, a list is made of all the distinct events that
might change the state of the system:
$4L-2$ events for $A$ and $B$ particle hops, $L$ events for
recombination of a pair, and $2$ events for insertion of a particle of
either type; altogether $5L$ events.
For each event $e_i$, a rate $r_i$ is calculated. Each step in the RED
simulation now consists of incrementing the time scale with
$\Delta t=1/\sum_j (r_j)$, and then allowing selection (and execution)
of an event. The probability that event $e_i$ is
chosen is equal to $p_i=r_i/\sum_j (r_j)$.

For an efficient implementation, a binary tree of events is
constructed, 
where each branch contains one event and has a weight equal to the
rate 
of that event. The weight of a parent node is equal to the sum of
the weights of its children. As the root node contains the sum of all
rates, the time increment $\Delta t$ is
easily obtained. For the selection of a particular event $e_i$
with rate $r_i$, 
we start in the root node, descend to one of its children with a
probability proportional to its weight, and iterate. The selected
event is then executed and the tree is updated.

The initial configuration for each simulation consisted of linear
density profiles for the A and B particles, which decreased from the
left and the right hand edges to the system center. In all our
simulations, we chose $L$ such that 
no A particle ever penetrated the B-rich region to within 10 sites of
the lattice boundary, and vice versa.
Correlation and thermalization times varied with $J$, $N$, $\lambda$
and $L$, and the tails of the reaction front required more
thermalization than the middle section. The necessary
thermalization time never exceeded $10^7$ events. To be safe, we
thermalized our system in all runs over $10^8$ events.

We consider first
the regime $\lambda J^{-1/2} D^{-1/2}\ll 1$, where the mean field
reaction front width $w_{mf}\sim(\lambda J/D^2)^{-1/3}$ is much larger 
than the predicted fluctuation modified width $\sim (J/D)^{-1/2}$. In
this case we expect 
that the behavior of the system should be close to mean field. 
Solving equations (\ref{mf1}) and (\ref{mf2}) for $a$ and $b$,
we find that the mean field reaction front $R_{mf}=\lambda
ab$ has the form
$R_{mf}=J(\lambda J/D^2)^{1/3}S([\lambda J/D^2]^{1/3}x)$,
where asymptotically (for $(\lambda J/D^2)^{1/3}|x|\gg 1$) we have
\begin{equation}
S\sim([\lambda J/D^2]^{1/3}|x|)^{3\over 4}\exp\left(
-{2\over 3}([\lambda J/D^2]^{1/3}|x|)^{3\over 2}\right)
\end{equation}

Hence the mean-field solution predicts that measured data for the
reaction front should
collapse if $R/[J(\lambda J/D^2)^{1/3}]$ is plotted as a function
of $(\lambda J/D^2)^{1/3}x$, as shown in figure 2.
In this case the number of particles $N$ (which varied from between
$1300$ to $12000$) was tuned to obtain the desired $J$. 

The collapsed data is in good agreement with the mean field
prediction, although there is a slight tendency for our simulation
data to lie to the right of $R_{mf}$, for the largest values of
$(\lambda J/D^2)^{1/3}x$.
In this region, where the number of minority particles is small, we
expect that noise from the reaction front will again become important,
leading to a widening of the profile. Note that these simulation
results were found not to depend on the inclusion of reservoirs in our
model, implying that the existence of correlations between particle
annihilation and reinjection was unimportant in this case.

In the limit $\lambda J^{-1/2}D^{-1/2}\gg 1$ the mean field solution
predicts that the reaction front will become increasingly
narrow. However, the
simulations disagree with this assertion - the reaction front
keeps a finite width even if $\lambda$ is made very large. Our
analysis, described below, distinguishes two components of this width:
one is intrinsic, and the other is caused by the
ability of the center of the front to wander. The intrinsic width is
calculable using the RG approach
already outlined, whereas the front wandering can be understood by
considering the fluctuations in the field
$\psi=a-b$, whose zero may be taken as defining the center of the
front. Including the effects of reaction front noise
(which is relevant in one dimension), the
field theory \cite{HC} leads to the following equation for $\psi$:
\begin{equation}
{\partial\psi\over\partial t}=D\nabla^2\psi+\eta. \label{noisydiff}
\end{equation}
Here $\eta$ is the reaction front noise, satisfying
$\langle\eta\rangle=0$ and
\begin{equation}
\langle\eta(x,t)\eta(x',t')\rangle=
2\delta(t-t')\delta(x-x')R,
\end{equation}
where the reaction rate at the wandering front, with width $w_g$, has
the form 
$R=(J/w_g)S(x/w_g)$. It is important to realize that, while $\psi$ is
on average equal to $\langle a\rangle -\langle b\rangle$, its
fluctuations are {\it not} the same as those in
the density difference. This arises from the non-trivial commutation
properties of the operators within the field theory.  More details on
this point (within the context of an $A+A\rightarrow\emptyset$
reaction) can be found in \cite{L}. This fact accounts for the
non-conservative nature of the noise in equation (\ref{noisydiff}).
We may now decompose $\psi$ into its mean field part together with
higher order Fourier harmonics:
\begin{equation}
\psi=-(J/D)x+\sum_{n=0}\chi_n(t)\cos\left({(2n+1)\pi x\over
L}\right)
\end{equation}
for $-(L/2)\leq x\leq (L/2)$. These corrections are the most general
possible which both couple to the noise (i.e. the harmonics have a
non-zero amplitude at $x=0$), and which are appropriate for the
non-conservative nature of the noise. Furthermore the
densities on the boundaries are kept constant by these additional
terms. We can now insert
the above expression into the noisy diffusion equation and Fourier
expand the reaction front noise (which is concentrated near $x=0$).
In the large time limit we find
\begin{equation}
\chi_n(t)\approx{2 \sqrt{2J} \over
L}\int_0^t\eta(t')\exp{\left[{(2n+1)^2\pi^2 D
\over L^2}(t'-t)\right]}dt',
\end{equation}
where now
\begin{equation}
\langle\eta(t)\rangle=0, \qquad
\langle\eta(t)\eta(t')\rangle=\delta(t-t').
\end{equation}
Clearly $\psi(x=0,t)=\sum\chi_n(t)$ is a Gaussian random variable
with
\begin{eqnarray}
& &\langle\psi(0,t)^2\rangle-\langle\psi(0,t)\rangle^2=\sum_{n,m}
\langle\chi_n(t)\chi_m(t)\rangle \\
& & \qquad\qquad\qquad\qquad\quad \sim{J\over \pi D}\ln{\left(cL\over
w_g\right)},
\end{eqnarray}
in the large time limit, with $c$ a constant, and where the upper
limit is provided by the finite width $w_g$ of the wandering reaction
front. Assuming that the 
fluctuations of $\psi$ are small in comparison with the system size,
then the gradient of $\psi$ at $x=0$ remains approximately equal to
$-(J/D)$. Hence 
to leading order we expect the position of the zero of the $\psi$
field to be a Gaussian random variable with width $w_g$, given by the
recursive relation:
\begin{equation}
w_g =\left[{\ln(cL/w_g)\over \pi
(J/D)}\right]^{1\over 2}. \label{wg}
\end{equation}
{}From our simulation data (in the limit $\lambda J^{-1/2} D^{-1/2}\gg
1$) we have plotted $R w_g/J$ as a function of
$x/w_g$, where in (\ref{wg}) we used $c=0.5$ (see figure 3).
The collapsed data is well
described by a normalized Gaussian, with width 1, in good
agreement with our theory. This indicates that the higher order
non-Gaussian corrections to the distribution of the zero of the $\psi$
field are indeed small. Note that the logarithmic factor in
(\ref{wg}) is essential for a good fit to the data. 

We can clearly see from figure 3
that the wandering Gaussian dominates over the intrinsic profile to
form the overwhelming component of the front.
Notice also that we find a basic $(J/D)^{-1/2}$ scaling, in
agreement with earlier predictions \cite{LC,CD,HC}. 
Previously, however, only the {\it intrinsic} part of the front was
being analyzed, whereas we have been studying the {\it wandering}
piece. The scaling agreement is simply a
consequence of dimensional analysis - any quantity with the dimensions
of length, which is
independent of $\lambda$, must scale as $(J/D)^{-1/2}$. 
We may also generalize our calculation to the time
dependent case, where a reaction front is formed quasistatically
between initially entirely segregated reactants. In this situation
we find $w_g\sim t^{\alpha}(\ln t)^{1/2}$, where $\alpha=1/4$. The
presence of the logarithm may explain the slow convergence found in
measurements of the exponent $\alpha$ \cite{C}. 

If we wish to study the intrinsic component of the front in the
non-mean 
field limit ($\lambda J^{-1/2}D^{-1/2}\gg 1$), we must now find a way of
suppressing the dominance of the wandering Gaussian part. One way
in which this can be achieved is to measure the
reaction rate $R_{r}$ as a function of $|x-x_{p}|$, the
distance between successive reaction events at $x_p$ and $x$. This
enables the 
intrinsic profile to be studied, as the front center has little time
to move on such short time scales. These reaction events
are effectively uncorrelated, so that the relative reaction
rate $R_{r}$ is given by
\begin{equation}
R_{r}(\tilde x)=\int R(x) P(x_p) \delta(x-x_p-\tilde x) dx dx_p,
\end{equation}
where $P(x_p)dx_p=R(x_p)dx_p/\int R(x_p)dx_p$ is the probability that
the previous reaction occurred between $x_p$ and $x_p+dx_p$.
In our simulations of $R_{r}$ we kept the insertion rate $i$
small, and hence many particles were present in the reservoirs.
We were therefore able to effectively simulate a much bigger
system, with a large number of particles. This ensured that
correlations between annihilation and reinsertion, which would
otherwise have modified the intrinsic profile, were kept to a minimum.
Our simulation data
is shown in figure 4, which shows a convincing data collapse of
$R_{r}(J/D)^{-1/2}/J$ plotted against
$|x-x_{p}|(J/D)^{1/2}$. As we are now studying the intrinsic profile,
this result finally confirms the scaling
predictions of \cite{LC,CD,HC}. For large values of
$|x-x_{p}|(J/D)^{1/2}$ we also find a tail which is consistent with the
RG improved tree level 
prediction $\log (R)\sim\ -B(J/D)^{1/2}|x|$ (implying $\log (R_r)\sim
-B(J/D)^{1/2}|\tilde x|$). Our data provides no clear indication of the
power law 
tails predicted in \cite{HC}. However, such tails would be very hard
to see in measurements of the relative reaction rate $R_r$, as
the power law exponent would be large ($7-2\epsilon+O(\epsilon^2)$).
Simulations have also been performed in the
mean field regime. In this case $R_{r}/[J(\lambda
J/D^2)^{1/3}]$ was found to collapse when plotted against small values
of $(\lambda J/D^2)^{1/3}|x-x_{p}|$. However, for larger values, the
collapse no longer 
worked well, probably due to the increased importance of fluctuations
in the asymptotic regime.

In summary, our ability to adjust the reaction rate has
enabled us to find two regimes for the
$A+B\rightarrow\emptyset$ front in one dimension. For $\lambda
J^{-1/2}D^{-1/2}\ll 1$ mean field predictions work well, whereas for
$\lambda 
J^{-1/2}D^{-1/2}\gg 1$ the front is dominated by a Gaussian profile, a result
of fluctuation induced wandering. Our theoretical prediction for this
shape agrees well with simulations. Finally, we have
succeeded in studying the intrinsic profile, where the shape and
scaling properties match previous RG calculations.

The authors thank S. Cornell and M. Droz for useful
discussions. We acknowledge financial support from the EPSRC,
under Grant No. GR/J78044 (GTB and JLC).

\bibliographystyle{prsty}

\begin{figure}
\label{model}
\caption{In our model, $A$ and $B$ particles are located on a
one-dimensional lattice. Three processes occur: particles can hop one
lattice site to the left or right with rate $h=1$; each pair of $A$
and $B$ particles at the same site can react with a rate
$\lambda$, after which 
they are moved to their respective reservoirs; and $A$ ($B$) particles
enter the lattice on the left (right) sides with an insertion rate
$i$.} 
\end{figure}

\begin{figure}
\caption{Collapsed data in the regime $\lambda J^{-1/2} D^{-1/2}\ll
1$. Solid line: mean field prediction. Squares: simulation results for
runs over $10^9$ events and $i=1000$, with $D=1$, $\lambda=0.001,
0.01, 0.1$, and $J=0.1, 0.2, 0.5, 1.0$.} 
\end{figure}

\begin{figure}
\caption{Collapsed data in the regime $\lambda J^{-1/2}D^{-1/2}\gg
1$. Solid 
line: normalized Gaussian; Data points: simulation results over $10^9$
events with $D=1$, $\lambda=1000$, $i=1000$, $J=0.1, 0.2, 0.5, 1.0$, and
$N=100$ (0), 1000 (+).} 
\end{figure}

\begin{figure}
\caption{Collapsed data for the relative reaction rate in
the regime $\lambda J^{-1/2}D^{-1/2}\gg 1$. Simulations were for $10^{10}$
events, with $D=1$, $N=200, 1100, 10100$.}
\end{figure}

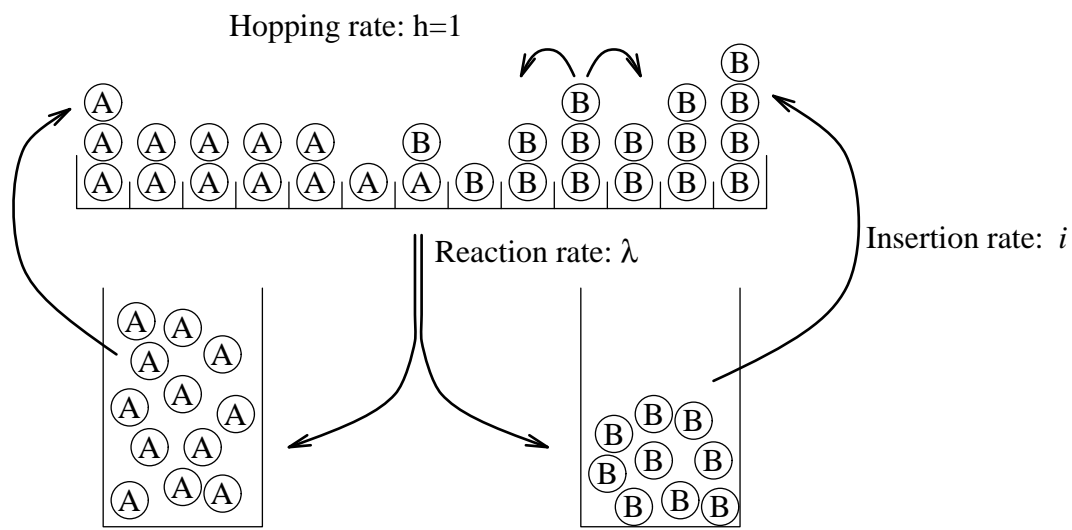
\end{document}